# Current-induced phase control in charged-ordered $Nd_{0.5}Ca_{0.5}MnO_3$ and $Pr_{0.6}Ca_{0.4}MnO_3$ crystals

**Sachin Parashar, L. Sudheendra, A.R. Raju and C.N.R. Rao**[*]

Chemistry and Physics of Materials Unit and CSIR Centre of Excellence, Jawaharlal Nehru Centre for Advanced Scientific Research, Jakkur. P.O., Bangalore-560064, India

**Abstract**

Single crystals of $Nd_{0.5}Ca_{0.5}MnO_3$ and $Pr_{0.6}Ca_{0.4}MnO_3$ show current-induced insulator-metal transitions at low temperatures. In addition, the charge-ordering transition temperature decreases with increasing current. The electroresistive ratio, defined as $\rho_{0.5}/\rho_I$ where $\rho_{0.5}$ is the resistivity at a current of 0.5 mA and $\rho_I$ the resistivity at a given applied current, I, varies markedly with temperature and the value of I. Thermal hysteresis observed in $Nd_{0.5}Ca_{0.5}MnO_3$ and $Pr_{0.6}Ca_{0.4}MnO_3$ at the insulator-metal transition indicates that the transition is first-order. The current-induced changes are comparable to those induced by magnetic fields, and the insulator-metal transition in $Pr_{0.6}Ca_{0.4}MnO_3$ is accordingly associated with a larger drop in resistivity.

PACS numbers: 71.30+h, 71.28+d, 72.40+w, 75.50-y

Corresponding Author: cnrrao@jncasr.ac.in    FAX # +91-80-8462760, +91-80-8462766

Phase control in manganates of the type $Ln_{1-x}A_xMnO_3$ (Ln=rare earth, A=alkaline earth) can be achieved by suitably choosing the Ln and A cations which control the bandwidth as well as the bandfilling. Thus, an appropriate A-site cations can lead to wide bandwidth conductors exhibiting ferromagnetism and metallicity through the double exchange mechanism. By reducing the bandwidth, charge ordering can be induced which owes its origin to coulomb repulsion, orbital ordering and Jahn-Teller distortion. The charge-ordered (CO) state is generally associated with antiferromagnetism and insulating behavior [1]. The control of phases in the manganates can lead to tunable electronic properties which is also achieved by subjecting them to external stimuli such as magnetic fields [2], irradiation by application of high power lasers [3] and chemical substitution [4,5]. Such external stimuli transform the CO state to a ferromagnetic metallic (FMM) state. The effects induced by the application of electric fields have been studied in the CO as well as the FMM states of the rare earth manganates [6-9]. The electric current drives these materials to a low resistive state, the conductive nature arising from the tunnel junctions separating neighboring ferromagnetic domains or due to the percolative nature of the conductive state. Current-induced memory effects and switching behavior have been reported for thin films and bulk samples of the manganates [8].

Since the effects of electrical fields on the CO rare earth manganates are somewhat unique, comparable to those of magnetic fields [1], we considered it important to investigate the effects of passing electrical currents through single-crystals of charge-ordered materials in greater detail. In particular we were interested to find out whether electric fields affect the CO transition, and whether the nature of the insulator-metal transition depends on the average size of the A-site cations. For this purpose, we have chosen $Nd_{0.5}Ca_{0.5}MnO_3$ and $Pr_{0.6}Ca_{0.4}MnO_3$, wherein the former with an average radius of the A-site cation, $<r_A>$, of 1.172 Å shows a robust CO state with a transition temperature, $T_{co}$, of 240 K which cannot be melted even under high magnetic fields of 25 T or greater. On the other hand, $Pr_{0.6}Ca_{0.4}MnO_3$ ($<r_A>$ = 1.18 Å) shows charge ordering around 230 K, and the CO state transforms to a FMM state at fields of 6-12 T. A comparison of the electric field effects on these two manganates is of interest, specially since it is known that an electric field-induced insulator-metal transition is accompanied by the appearance of magnetization [10].

Single crystals of $Nd_{0.5}Ca_{0.5}MnO_3$ and $Pr_{0.6}Ca_{0.4}MnO_3$ were grown by the floating-zone furnace fitted with two ellipsoid halogen lamps, having prepared the polycrystalline samples of the materials by the solid state route. Monophasic polycrystalline samples were hydrostatically pressed and sintered at 1400 $^o$C for 24 hrs to obtain feed and seed rods of dimensions 8cm in length and 4mm in diameter. Single crystals were then grown under 2-3 l/min of airflow. The crystals thus obtained was cut and subjected to oxygen annealing for 48h. Magnetization data were obtained with a vibrating sample magnetometer operating between 300 - 50 K. Electrical resistivity measurements were carried out on crystal of dimensions 4 mm in diameter and 1 mm in thickness by the standard four-probe method. The resistivity was measured two ways. 1) The sample was cooled under a constant current and upon reaching the lowest temperature, data were collected in the warming cycle without switching off the current. 2) The sample was cooled under a constant current to the lowest possible temperature and the current turned off for before starting the warming up cycle under the same constant current.

In fig.1a, we show the resistivity of a $Nd_{0.5}Ca_{0.5}MnO_3$ crystal at different applied currents when the sample is cooled from 300 K. There are four distinct features in the plot. There is a drop in the resistivity throughout the temperature range as the current, I, is increased. The temperature dependence of the resistivity changes with the increase in I. An insulator-metal transition occurs around 65 K ($T_{IM}$) at high values of I, beyond a threshold value. The $T_{co}$ shifts to lower values with increase in I and become constant beyond a high I value (~50mA). Such a current-induced shift in $T_{co}$ is indeed noteworthy. The plot of $T_{co}$ against I (see the inset of Fig. 1a) is linear with a slope of ~1 K/mA. The decrease in the $T_{co}$ could be due to the charge delocalization driven by external current which in turn decreases lattice distortions.

As the current through the $Nd_{0.5}Ca_{0.5}MnO_3$ crystal is increased, there is a drastic change in the resistivity. The $\rho_{0.5}/\rho_I$ ratio where $\rho_{0.5}$ is the resistivity at a current of 0.5 mA (which is the smallest I employed by us) and $\rho_I$ the resistivity at a given applied current, I, may be considered to represent the electroresistive ratio analogous to the magnetoresistive ratio. In Fig.1b we show the temperature variation of $\rho_{0.5}/\rho_I$ for different I values. The change in resistivity at low temperatures (~20 K) is four orders of magnitude while it is one or two orders at high temperatures (T ≈ $T_{co}$). The four orders of magnitude change in the resistivity is achieved by changing the applied current by two orders, indicating the non-ohmic behavior. As the current is increased, we observe a sharp drop in resistivity of the sample at the insulator-metal transition temperature, $T_{IM}$ (Fig. 2a). a behavior not noticed in earlier films.

When the ρ-T data are recorded in the warming run without turning off the current, we observe a thermal hysteresis, which becomes prominent at large currents. The observation of hysteresis suggests a first-order nature of the structural change induced in the transition. In the warming run, the low resistive state persists down to a temperature $T_W$, lower than $T_{IM}$ (fig.2a). The resistivity change at $T_W$ increases with the increasing current. Both $T_{IM}$ and $T_W$ are independent of the applied current as also the ratio of the peak resistivity in the heating and the cooling runs. If one compares the change in the resistivity to the ratio of resistivities, the phenomenon is quite striking (fig.2b).

When the current is switched off at low temperatures, the current source is not able to pass the same current in the sample when the current is again switched on at the same low temperature due to voltage limit (V-Limit 105 V) of the source. This implies that the resistivity of the sample becomes high on switching off the current and the source meter is unable to pass the required current with the maximum voltage drop it can generate across the leads (which was possible before switching off the current). The effect of the current in the warming run is quite different from that in the cooling run, the resistivities in the former being higher. Although the resistivity of the sample has field dependence, all the $\rho$-T curves merge around 150 K above which the current has little effect (fig.2a). This is in stark contrast to the cooling run data.

Electric-current-induced effects on a $Pr_{0.6}Ca_{0.4}MnO_3$ crystal are shown in fig.3. The effects are qualitatively similar to those in $Nd_{0.5}Ca_{0.5}MnO_3$, but the drop in resistivity at low temperatures is considerably more marked. This is noteworthy feature as discussed later. The insulator-metal transition observed in the charge-ordered crystals is not due to Joule heating, as the crystals kept at 20 K under different currents for a period of 1-2 h did not show any change in resistivity although the resistivity decreased with increasing current.

The main aspects of the present study are as follows. Passing relatively high currents through the charge-ordered manganates, as a general rule, cause an insulator-metal transition at $T_{IM}$. This is unlike the effect of magnetic field in the charge-ordered manganates where only those with $<r_A> > 1.18$ Å could be transformed into the metallic state at fields $\leq 15$ T. Accordingly, both $Nd_{0.5}Ca_{0.5}MnO_3$ and $Pr_{0.6}Ca_{0.4}MnO_3$ attain the metallic state at low temperatures and high currents. The electroresistive ratio increases with increasing current and the charge ordering transition temperature decreases with increasing current. The magnitude of drop in the resistivity at the insulator-metal transition parallels the effect of magnetic fields in some respect. Thus, the drop in resistivity in $Pr_{0.6}Ca_{0.4}MnO_3$ is much higher than that of $Nd_{0.5}Ca_{0.5}MnO_3$ for similar currents. It may be noted that $Nd_{0.5}Ca_{0.5}MnO_3$ contains two phases below $T_{co}$ [11], and ferromagnetic correlations are present at low temperatures [1,12]. The presence of two phases below $T_{co}$, in $Pr_{0.6}Ca_{0.4}MnO_3$ with one phase having spin-glass like features has also been shown recently [13]. On the application of a reasonably high electric current, metallic domains are likely to grow in size. Such a metal-like phase generated by electric fields will also be magnetic [10]. The metal-like conduction around 65 K may be facilitated by the separation of FMM, CO insulator and paramagnetic phases.

**Acknowledgement**

The authors thank BRNS (DAE) and DRDO (India) for support of this research.

**Figure Captions**

Fig.1 (a) Temperature variation of resistivity of $Nd_{0.5}Ca_{0.5}MnO_3$ in cooling runs. Inset shows the variation of $T_{co}$ with the applied current. (b) Variation of the electroresistive ratio with temperature in the cooling cycle. Inset shows the same in the heating cycle.

Fig. 2 (a) Temperature variation of resistivity of $Nd_{0.5}Ca_{0.5}MnO_3$ in the cooling and heating cycles, wherein the current is not switched off at the lowest temperature before starting the heating run. The inset shows the same with the current switched off and then switched on at the lowest temperature, before starting the heating run. (b) The squares represent the ratio of resistivities at the peak value in the cooling and heating runs with the current not switched off before starting measurements at the lowest temperature. Open circles represent the difference in the peak values of resistivity.

Fig. 3 Temperature variation of resistivity of $Pr_{0.6}Ca_{0.4}MnO_3$ at different currents

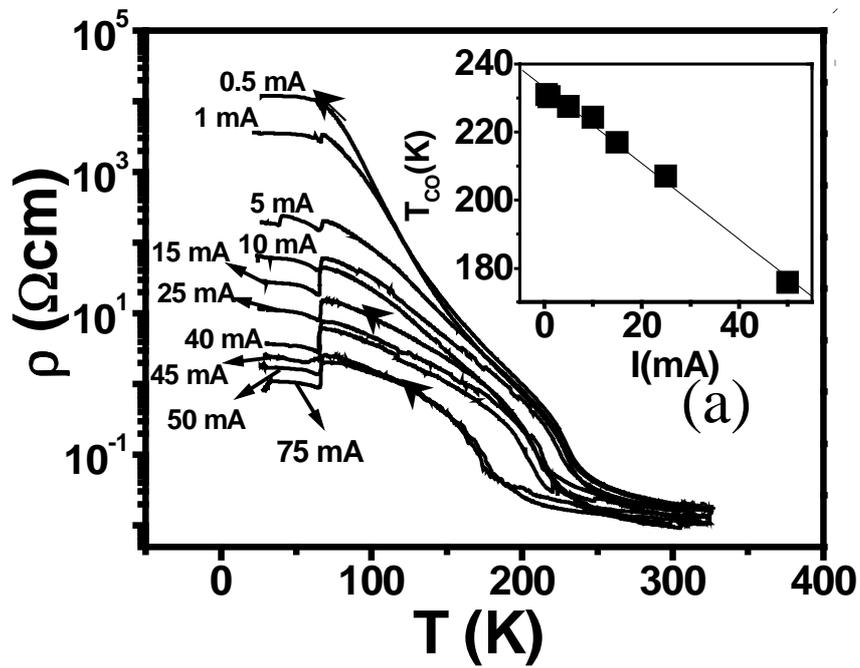

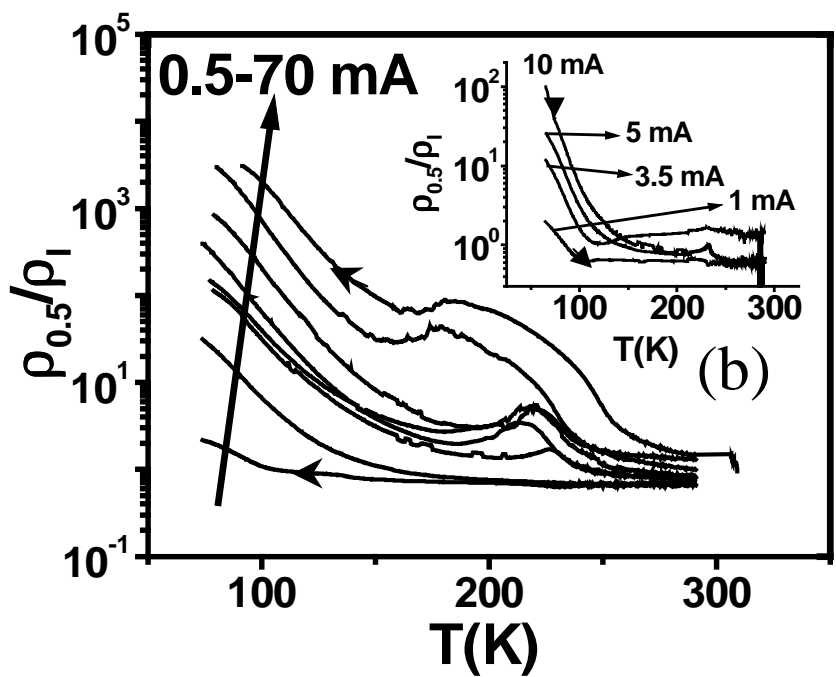

Parashar et.al
Fig. 1

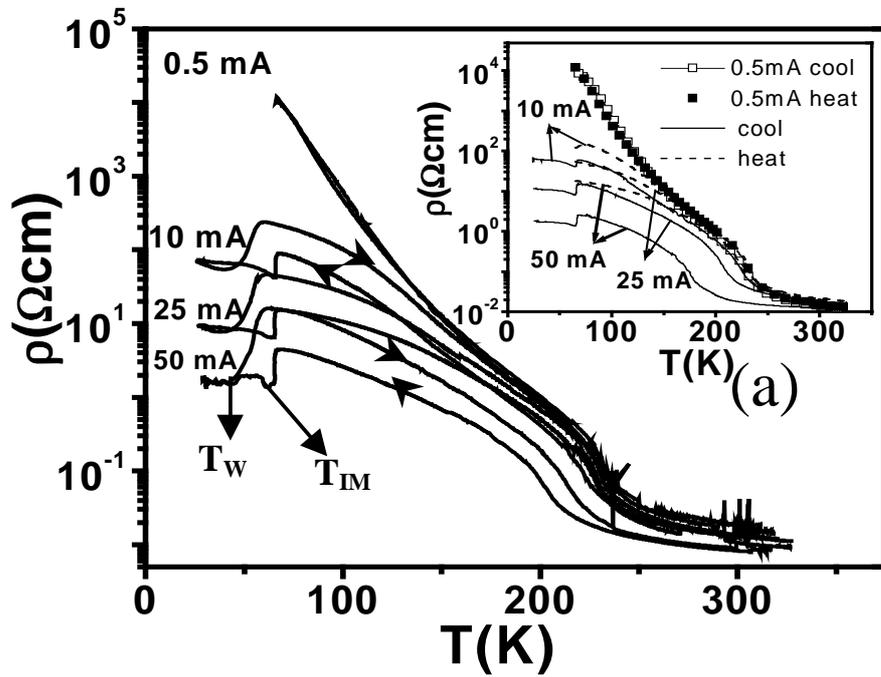
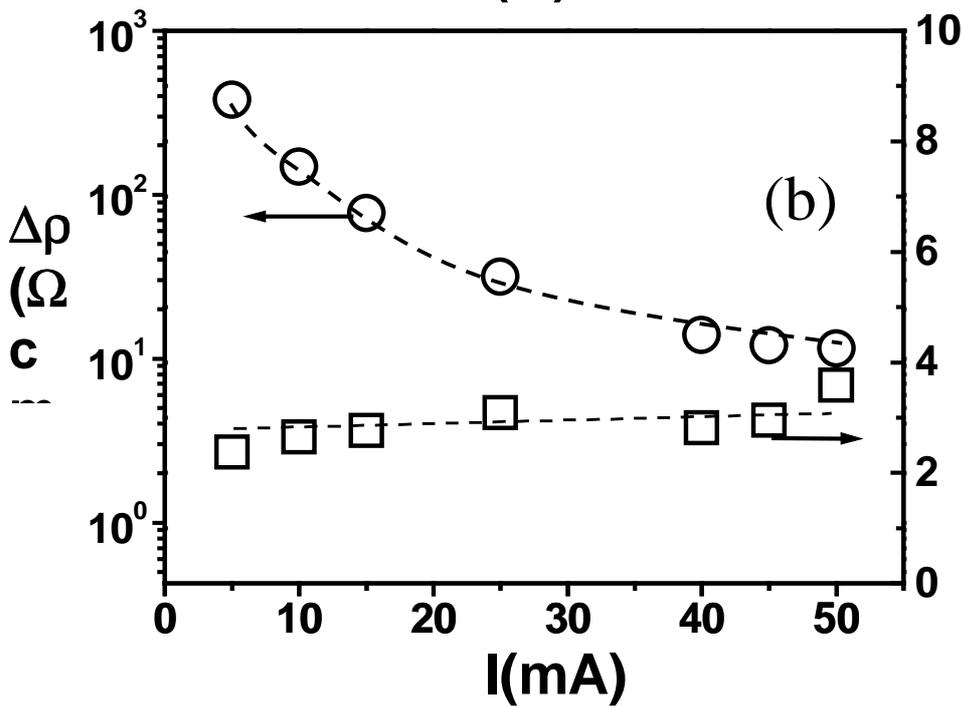

Parashar et.al
Fig. 2

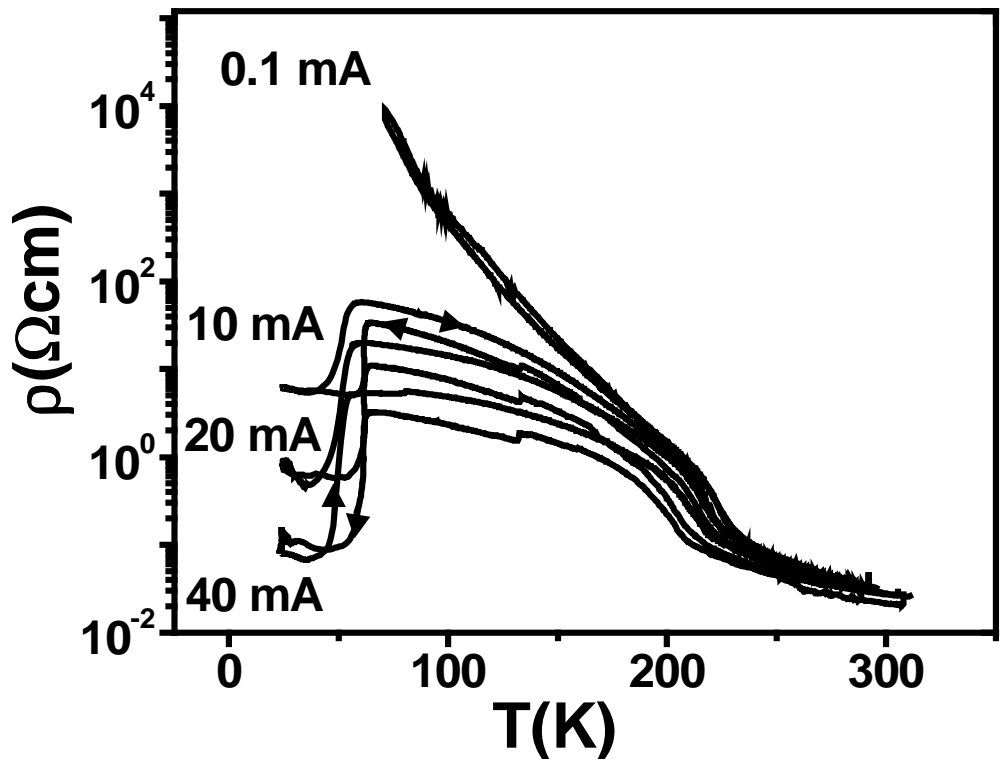

**Parashar et. al
Fig. 3**